\documentclass[aps,prd,twocolumn,nofootinbib,eqsecnum]{revtex4}
\usepackage{verbatim}
\usepackage{amssymb}
\usepackage{amsmath}
\usepackage{amsfonts}
\usepackage{bm}
\usepackage{hyperref}
\usepackage{graphicx}
\usepackage{tensor}
\usepackage{mathrsfs}

\usepackage{enumitem}

\newcommand{\doe}{\partial}
\newcommand{\be}{\begin{equation}}
\newcommand{\ee}{\end{equation}}
\newcommand{\bea}{\begin{eqnarray}}
\newcommand{\eea}{\end{eqnarray}}
\newcommand{\bdm}{\begin{displaymath}}
\newcommand{\edm}{\end{displaymath}}
\newcommand{\bse}{\begin{subequations}}
\newcommand{\ese}{\end{subequations}}

\newcommand{\mr}{\mathrm}

\newcommand{\mc}{\mathcal}

\newcommand{\py}{\phantom{yo}}
\newcommand{\qq}{\quad\quad}

\begin{document}

\title{Is motion under the conservative self-force in black hole spacetimes\\ an integrable Hamiltonian system?}

\author{Justin Vines and \'Eanna \'E. Flanagan}
\affiliation{Department of Physics, Cornell University, Ithaca, NY 14853, USA.}

\begin{abstract}

A point-like object moving in a background black hole spacetime experiences a gravitational self-force which
can be expressed as a local function of the object's instantaneous position and
velocity, to linear order in the mass ratio.
We consider the worldline dynamics defined by the conservative part of the local self-force, turning
off the dissipative part, and we ask:  Is that
dynamical system a Hamiltonian system, and if so, is it integrable?

In the Schwarzschild spacetime, we show that the system is Hamiltonian
and integrable, to linear order in the mass ratio, for generic (but not necessarily all)
stable bound orbits.  There exist an energy and an angular
momentum, being perturbed versions of their counterparts for geodesic
motion, which are conserved under the forced motion.  We also
discuss difficulties associated with establishing analogous results in
the Kerr spacetime.
This result may be useful for future computational schemes, based on a local Hamiltonian
description, for calculating the conservative
self-force and its observable effects.  It is also relevant
to the assumption of the existence of a Hamiltonian for the
conservative dynamics for generic orbits in the effective-one-body formalism, to linear
order in the mass ratio, but to all orders in the post-Newtonian
expansion.

\end{abstract}

\maketitle

\section{Introduction}\label{sec:intro}

Recent years have seen great progress in computing the motion of and
gravitational wave signals emitted by point-like objects
orbiting massive black holes
\cite{Detweiler_rev05,MPP,Barack_rev,Barack_rev14}, but some key
foundational and computational challenges still lie ahead
\cite{Barack_rev14}.  A sufficiently accurate solution to the
gravitational self-force problem will be crucial to the detection and
analysis of signals from astrophysical extreme-mass-ratio inspirals by
future space-based gravitational wave detectors \cite{lisa}.  Study
of the self-force also helps to inform other complementary approaches to the
relativistic two-body problem, such as post-Newtonian theory and the
effective-one-body formalism \cite{LeTiec_rev}.



The foundations for computing the first-order self-force acting on a
point mass in a curved background spacetime are provided by a result
\cite{MiSaTa,QuWa} which has been canonized as the `MiSaTaQuWa
equation'.  This is an equation for the self-force experienced by the point
mass as a functional of its worldline history, arising from the tail
part of its own linearized gravitational field.
There are several different formulations of the \mbox{MiSaTaQuWa}
result:
\begin{enumerate}
\item \emph{A correction to the geodesic deviation equation:}\label{gde}
In this formulation, the deviation between the point mass's actual
accelerated worldline and a given fiducial background geodesic is
characterized by a deviation vector field along the fiducial geodesic.
The deviation vector obeys the geodesic deviation equation plus a
self-force correction, and the self-force correction is constructed
from the self-field sourced by the fiducial geodesic.  This
formulation is the version that is obtained directly from the most
rigorous derivations \cite{GrallaWaldDGSF,rigor}.

\item \emph{A local equation of motion for an accelerated worldline:}
In this scheme, the worldline's acceleration is given as a function of
its instantaneous position and velocity, by using the geodesic which
instantaneously matches the position and velocity of the accelerated
worldline (the osculating geodesic) as the source to compute
the instantaneous self-force \cite{MPP}.

\item \emph{A non-local equation of motion for an accelerated worldline:}
In this family of schemes, one uses the history of the actual
accelerated worldline as the source for the self-force\footnote{Since
  the stress energy tensor for an accelerated worldline is not
  conserved, implementing this sheme requires an ad hoc relaxation of
  the Lorentz gauge condition.}.  Viewing the
self-force as a given functional of the worldline, this corresponds to
a non-local equation of motion for the the worldline---the
instantaneous acceleration depends on the worldline's entire history.

\end{enumerate}

Schemes 2 and 3 have a clear advantage over scheme 1 in being able to
describe inspiraling motion over a radiation
reaction timescale.  Both 2 and 3 reduce to 1
for small deviations from the fiducial geodesic, over short timescales.
Scheme 2 is said to arise from 3 via a reduction of order
procedure: because the self-forced worldline and the osculating
geodesic agree to zeroth order in the mass ratio, the former can be
replaced with the latter in calculating the self-force to first order.
The relative accuracies and computational merits of 2 and variants of
3 are subtle issues, entangled with formulating and implementing a
consistent second-order analog of the MiSaTaQuWa result
\cite{Rosenthal_second-order,Gralla_second-order,Pound_second-order,Pound_nonlinear},
which will not be addressed here.

This paper addresses a formal question concerning the worldline
dynamics
 defined by scheme 2, using
only the conservative part of the self-force, turning off the
dissipative/radiative part.  We ask:  Is the resultant dynamical
system, for the worldline degrees of freedom only, an integrable
Hamiltonian system, to linear order in the mass ratio?

In the case where the background spacetime is Schwarzschild, we show
that this system is
indeed a Hamiltonian dynamical system, in the classic sense
\cite{Arnold}.  We also show that it is
integrable; in other words, there exist three functions (the rest mass, an
energy, and an angular momentum) on the effectively six-dimensional
phase space whose values are instantaneously conserved along the
system's trajectories.
These results hold to linear order in the mass
ratio, for generic stable bound orbits (with some caveats for orbits
in the zoom-whirl regime).

We also carry out a partial analysis of the case of the Kerr
spacetime, following the same line of reasoning which leads to a
successful construction in Schwarzschild.  In Kerr, we encounter
complications which are associated with orbital resonances
\cite{resonances}, when the ratio of the frequencies of the radial and
polar motions is a rational number.
These complications prevent us from
drawing definite conclusions about the Kerr case.

There is, however, a heuristic argument which suggests that the motion
in Kerr should be integrable and Hamiltonian
\cite{two_timescale,Wu10}.  Namely, the ambiguities in the definition
of angular momentum related to the BMS group are associated with the
dissipative sector of the linear perturbations, so the conservative
sector should admit three independent conserved components of angular
momentum \cite{wolfgang}, a sufficient number to make the system
integrable.  This
general picture is consistent with what has been found in the
post-Newtonian approximation, where the system is integrable at
successive orders \cite{Wu10}.

The conservative self-force dynamics in Kerr has recently been cast in
a Hamiltonian-like formulation in Ref.~\cite{Isoyama14}.  However,
that formulation is not Hamiltonian in the classic sense used here.
Specifically, in that formalism, the equations of motion are obtained
by differentiating the Hamiltonian with respect to one set of
variables while holding a second set of variables fixed, and then
taking the two sets of variables to coincide.

The existence of a Hamiltonian system for the conservative dynamics is a foundational assumption in the effective-one-body formalism \cite{intro_eob}, as well as in other related treatments of the relativistic two-body problem, e.g.~\cite{Buonanno03,LeTiec12,binding_energy}.  While the results of Refs.~\cite{Barausse11,LeTiec12,binding_energy} make it clear that such a Hamiltonian exists for circular orbits in the extreme-mass-ratio limit (to linear order in the mass ratio, but to all orders in the post-Newtonian expansion), our result confirms the validity of this assumption for generic (non-circular) orbits, in the non-spinning case.

\py

This paper is organized as follows.  Our construction relies heavily
on the fact that geodesic motion in the Schwarzschild (and Kerr)
spacetime is completely integrable, and thus admits a representation
in terms of generalized action-angle variables
\cite{Schmidt,two_timescale}.  We review relevant properties of these
variables in Sec.~\ref{sec:geod}, and we review relevant properties of
the local conservative self-force in Sec.~\ref{sec:force}.  We develop
sufficient conditions for the forced system to be Hamiltonian and
integrable in Sec.~\ref{sec:question}, and we conclude in
Sec.~\ref{sec:conclude}.




\section{geodesic motion in Kerr and action-angle variables}\label{sec:geod}

Geodesic motion in a spacetime with metric $g_{\mu\nu}$ is generated by the Hamiltonian
\be
H_0(z,p)=\frac{1}{2}g^{\mu\nu}(z)p_\mu p_\nu,
\ee
where $z^\mu(\lambda)$ are the coordinates of a worldline,
and $p_\mu(\lambda)$ are the components of its momentum, together with the
canonical symplectic form $\Omega_0=\mr dz^\mu\wedge \mr dp_\mu$ on
the worldline phase space $(z^\mu,p_\mu)$ \cite{Arnold,two_timescale}.
Hamilton's equations read
\be\label{geod_notcov}
\frac{dz^\mu}{d\lambda}=\frac{\doe H_0}{\doe p_\mu},\qq \frac{dp_\mu}{d\lambda}=-\frac{\doe H_0}{\doe z^\mu},
\ee
and imply the geodesic equation in affine parameterization,
$p^\nu\nabla_\nu p_\mu=0$.  Identifying the affine parameter as
$\lambda=\tau/m$, where $\tau$ is the proper time along the worldline,
gives the usual expression $p_\mu=mg_{\mu\nu}dz^\nu/d\tau$ for the
momentum of a particle of mass $m$, and also yields $H_0=-m^2/2$ on
shell.

In the Kerr spacetime, with Boyer-Lindquist coordinates
$z^\mu=(t,r,\theta,\phi)$, geodesic motion is completely integrable
thanks to the existence of four first integrals of motion.  These are
the Hamiltonian
$H_0$, the energy $E=-(\doe_t)^\mu p_\mu$, the axial angular momentum
$L_z=(\doe_\phi)^\mu p_\mu$, and the Carter constant
$Q=K^{\mu\nu}p_\mu p_\nu$, where $(\doe_t)^\mu$ and $(\doe_\phi)^\mu$
are the timelike and axial Killing vectors, and $K^{\mu\nu}$ is the
non-trivial Killing tensor.

These four first integrals are independent and in involution, which
implies the existence of
generalized action-angle variables\footnote{The action angle variables discussed here
are associated with the affine parameter $\lambda=\tau/m$ along
the worldline; there are also other action-angle coordinates on the
phase space associated with Mino time \cite{Mino} and Boyer-Lindquist coordinate time.}
for bound geodesics in the Kerr
geometry \cite{Schmidt,two_timescale}.  These are canonical coordinates
$\big(q^\alpha,J_\alpha\big)=(q^t,q^r,q^\theta,q^\phi,J_t,J_r,J_\theta,J_\phi)$
on the worldline phase space, for which
the geodesic Hamiltonian $H_0$ depends only on the action variables
$J_\alpha$,
\be\label{HJ}
H_0(z,p)=H_0(J),
\ee
and not on the angle variables $q^\alpha$.  They are
obtained from the $(z^\mu,p_\mu)$
coordinates via a canonical transformation
\be
\label{cantra}
q^\alpha = q^\alpha(z^\mu,p_\mu),\ \ \ \
J_\alpha = J_\alpha(z^\mu,p_\mu),
\ee
so that $\Omega_0=\mr dz^\mu\wedge \mr dp_\mu=\mr dq_\alpha\wedge\mr dJ_\alpha$.
Hamilton's equations then take the particularly simple form
\bse
\label{HamJ}
\bea
\frac{dq^\alpha}{d\lambda}&=&\frac{\doe H_0}{\doe
  J_\alpha}\equiv\omega^\alpha(J), \\
\frac{dJ_\alpha}{d\lambda}&=&-\frac{\doe H_0}{\doe q^\alpha}=0.
\eea
\ese
The action variables $J_\alpha$ are all independent constants of
motion, and the angle variables $q^\alpha$ all increase linearly, at
the constant rates $\omega^\alpha$ known as the fundamental
frequencies.  The angle variables $q^r$, $q^\theta$, and $q^\phi$ are
each periodic with period $2\pi$, while $q^t$ has an infinite range.
The action variables $J_\alpha$ are functions of the geodesic first
integrals $P_\alpha\equiv(H_0,E,L_z,Q)$; in particular, $J_t=-E$ and $J_\phi=L_z$.

In the Schwarzschild limit of the Kerr geometry, both geodesic motion
and self-forced motion are
confined to a plane, which can be taken without loss of generality to
be the equatorial plane $\theta=\pi/2$.  We can then ignore the
$\theta$-motion, working in the reduced phase space with coordinates
$(z^\mu,p_\mu)=(t,r,\phi,p_t,p_r,p_\phi)$.  We have the three first
integrals $P_\alpha=(H_0,E,L_z)$, and action-angle variables
$\big(q^\alpha,J_\alpha\big)=(q^t,q^r,q^\phi,J_t,J_r,J_\phi)$, defined
just as in the general Kerr case.

\section{conservative-self-force perturbation to geodesic motion}\label{sec:force}

Instead of geodesic motion, we now consider a point mass $m$
with worldline $z^\mu(\lambda)$ experiencing a local linear perturbing
force,
\be\label{zpforced}
\frac{dz^\mu}{d\lambda}=p^\mu,\qq p^\nu\nabla_\nu p_\mu=\epsilon \mc F_\mu(z,p),
\ee
where $\epsilon$ is a small parameter.  We are interested in the case
where the forcing function $\mc F_\mu(z,p)$ is given by the
conservative part of the osculating-geodesic-sourced first-order
gravitational self-force in the Kerr spacetime, with the parameter
$\epsilon$ being the small mass ratio $m/M$.

The worldline dynamics (\ref{zpforced})
can be re-expressed in terms of the geodesic action-angle variables $(q^\alpha,J_\alpha)$ as
\bse
\label{qJsystem}
\bea
\frac{dq^\alpha}{d\lambda}&=&\omega^\alpha(J)+\epsilon
f^\alpha(q,J), \\
\frac{dJ_\alpha}{d\lambda}&=&\epsilon F_\alpha(q,J),
\eea
\ese
following Ref.~\cite{two_timescale}.  Here we use
the same phase space coordinate transformation (\ref{cantra})
as for geodesic motion
to obtain Eqs.\ (\ref{qJsystem}) from Eqs.\ (\ref{zpforced}).
The forcing functions $f^\alpha(q,J)$ and
$F_\alpha(q,J)$ are determined from the self-force components $\mc
F_\mu(z,p)$ via $f^\alpha=({\doe q^\alpha}/{\doe p_\mu})_z\,\mc F_\mu$
and $F_\alpha=({\doe J_\alpha}/{\doe p_\mu})_z\,\mc F_\mu$.
These functions have the important property that they are independent
of the angle variables $q^t$ and $q^\phi$, because of the symmetries
of the Kerr spacetime.  Thus, they can written as functions of $q^r$,
$q^\theta$ and the four variables $J_\alpha$ \cite{two_timescale}:
\bse
\label{fFalpha_ind}
\bea
f^\alpha(q,J)&=&f^\alpha\big(q^r,q^\theta,J\big), \\
 F_\alpha(q,J)&=&F_\alpha\big(q^r,q^\theta,J\big).
\eea
\ese
In the Schwarzschild case, these functions depend only on $q^r$ and
the three $J_\alpha$.

The forcing functions $f^\alpha$ and $F_\alpha$ are periodic
functions of $q^r$ and of $q^\theta$, each with period $2\pi$.  They
can thus be expanded as Fourier series in $q^r$ and $q^\theta$,
according to
\be\label{Fouri}
F_\alpha\big(q^r,q^\theta,J\big)=\sum_{k_r,k_\theta=-\infty}^\infty\;
{\hat F_\alpha}(k_r, k_\theta,J)\;e^{ik_r q^r+ik_\theta q^\theta},
\ee
and similarly for $f^\alpha$.
As shown by Mino \cite{Mino},
the $(0,0)$ Fourier mode of each $F_\alpha$ vanishes,
\be\label{zeromode}
\hat F_\alpha(0,0,J)=\frac{1}{(2\pi)^2}\int_0^{2\pi}dq^r\int_0^{2\pi}dq^\theta\,F_\alpha(q^r,q^\theta,J)=0,
\ee
due to reflection properties of Kerr geodesics and to the time-reversal symmetry of the conservative self-force \cite{two_timescale,Mino}. 

In the Schwarzschild case, we lose the dependence on the
$\theta$-motion.  Equations (\ref{fFalpha_ind}) and (\ref{Fouri}) are replaced by
\be\label{FouriSchw}
F_\alpha\big(q^r,J\big)=\sum_{k_r=-\infty}^\infty
{\hat F_\alpha}(k_r,J)\;e^{ik_r q^r},
\ee
and similarly with $F_\alpha$ replaced by $f^\alpha$.
Equation (\ref{zeromode}) becomes
\be\label{zeromodeSchw}
\hat F_\alpha(0,J)=\frac{1}{2\pi}\int_0^{2\pi}dq^r\,F_\alpha(q^r,J)=0.
\ee

Recalling that the forcing functions $F_\alpha$ give the rates of change of the action
variables $J_\alpha$, Eqs.~(\ref{zeromode}) and (\ref{zeromodeSchw})
express the fact that the conservative first-order self-force causes no net change
in the geodesic first integrals, when the force is evaluated along a
geodesic and suitably averaged \cite{Mino}.  In
the Schwarzschild case (\ref{zeromodeSchw}), the averaging is
an orbital average or time average over one period of radial motion.
In the Kerr case (\ref{zeromode}), the average is over the
$(q^r,q^\theta)$ torus in phase space, which is equivalent to a time average
only over an infinite time and only for non-resonant orbits
\cite{two_timescale,resonances,Grossman:2011im}.


\section{Is the perturbed system Hamiltonian and integrable?}\label{sec:question}

The perturbed system (\ref{qJsystem}) will be Hamiltonian
and integrable, to linear order in $\epsilon$, if
there exist new phase space coordinates $(\bar q^\alpha,\bar
J_\alpha)$ and a new Hamiltonian function ${\bar H}({\bar J})$
for which Eqs.~(\ref{qJsystem}) are equivalent to
\be\label{qJbarsystem}
\frac{d\bar q^\alpha}{d\lambda}=\frac{\doe \bar H(\bar J)}{\doe \bar J_\alpha}+O(\epsilon^2),\qq
\frac{d\bar J_\alpha}{d\lambda}=O(\epsilon^2).
\ee
%
%
Without loss of generality,
we can express the new coordinates as linear perturbations of the
geodesic action-angle coordinates $(q^\alpha,J_\alpha)$:
\bse
\label{qJbar}
\bea
\bar q^\alpha(q,J)&=&q^\alpha+\epsilon\chi^\alpha(q,J), \\
 \bar J_\alpha(q,J)&=&J_\alpha+\epsilon \zeta_\alpha(q,J),
\eea
\ese
for some functions $\chi^\alpha$ and $\zeta_\alpha$ to be determined.
Note that (\ref{qJbar}) is not assumed to be a canonical transformation.
Similarly we can express the new Hamiltonian as
\be\label{Hbar}
\bar H(\bar J)=H_0(\bar J)+\epsilon H_1(\bar J),
\ee
where $H_0$ is the geodesic Hamiltonian function, for some function
$H_1(\bar J)$ to be determined.

Combining Eqs.\ (\ref{qJbarsystem}), (\ref{qJbar}) and (\ref{Hbar})
now yields that
Eqs.~(\ref{qJbarsystem}) will be equivalent to Eqs.~(\ref{qJsystem}) to linear order in $\epsilon$ if
\bse
\bea
\label{fcond}
f^\alpha(q,J)&=&-\omega^\beta \frac{\doe\chi^\alpha}{\doe
  q^\beta}+\frac{\doe\omega^\alpha}{\doe
  J_\beta}\zeta_\beta+\frac{\doe H_1(J)}{\doe J_\alpha},\ \ \
\\\label{Fcond}
F_\alpha(q,J)&=&-\omega^\beta\frac{\doe \zeta_\alpha}{\doe q^\beta},
\eea
\ese
where $\omega^\alpha=\omega^\alpha(J)$ are the geodesic fundamental
frequencies (\ref{HamJ}).  Thus, the conservative-self-force
dynamics (\ref{qJsystem}) will be Hamiltonian and integrable if there
exist functions $\chi^\alpha(q,J)$, $\zeta_\alpha(q,J)$, and $H_1(J)$
satisfying Eqs.~(\ref{fcond}) and (\ref{Fcond}).

In light of Eqs.~(\ref{fFalpha_ind}), it is natural to consider solutions for $\chi^\alpha$ and $\zeta_\alpha$ which, like $f^\alpha$ and $F_\alpha$, are independent of $q^t$ and $q^\phi$, and which are periodic functions of $q^r$ and $q^\theta$ [or of just $q^r$ in Schwarzschild].  We can then decompose all of these functions into discrete Fourier series for the $q^r$ and $q^\theta$ dependence, just as for $F_\alpha$ in Eq.~(\ref{Fouri}) [or Eq.~(\ref{FouriSchw})].  This defines Fourier mode amplitudes $\hat f^\alpha$, $\hat F_\alpha$, $\hat \chi^\alpha$, and $\hat\zeta_\alpha$ which are functions of the two integers $k_r$ and $k_\theta$ [or just $k_r$] and all of the action variables $J_\alpha$.

We then have the following Fourier transforms of Eqs.~(\ref{fcond}) and (\ref{Fcond}):
\bse
\bea
\label{fcondFouri}
\hat f^\alpha&=&-i(\omega\cdot
k)\hat\chi^\alpha+\frac{\doe\omega^\alpha}{\doe
  J_\beta}\hat\zeta_\beta+\delta_{k_r,0}\delta_{k_\theta,0}\frac{\doe
  H_1}{\doe J_\alpha}, \ \ \ \
\\\label{FcondFouri}
\hat F_\alpha&=&-i(\omega\cdot k)\hat \zeta_\alpha,
\eea
\ese
where
\be
(\omega\cdot k)=\left\{\begin{array}{ll}\omega^r k_r+\omega^\theta k_\theta\qq&\mr{Kerr}
\\
\omega^r k_r&\mr{Schwarzschild}
\end{array}\right.,
\ee
and with $\delta_{k_\theta,0}\to 1$ in Schwarzschild.
If we restrict attention to Fourier modes for which $(\omega\cdot
k)\ne 0$, then Eqs.~(\ref{fcondFouri}) and (\ref{FcondFouri}) admit
the simple solutions
\be\label{zeta_sol}
\hat\zeta_\alpha=\frac{i\hat F_\alpha}{(\omega\cdot k)}, \qq\hat\chi^\alpha=\frac{i}{(\omega\cdot k)}\left(\hat f^\alpha- \frac{\doe\omega^\alpha}{\doe J_\beta}\hat\zeta_\beta\right).
\ee

In the general Kerr case, the quantity $\omega\cdot k=\omega^r
k_r+\omega^\theta k_\theta$ can vanish at locations in phase space
where $\omega^r/\omega^\theta$ is a rational number, corresponding to
an orbital resonance in the $r$ and $\theta$ motions
\cite{resonances}.  The solutions (\ref{zeta_sol}) are clearly not
valid in such cases, and so
our analysis
does not allow us to draw any definite conclusions about
the Kerr case.

For stable bound orbits in Schwarzschild, the quantity $\omega\cdot
k=\omega^r k_r$ vanishes only when $k_r=0$, since $\omega^r=0$ occurs
only in the limit of unbound or unstable orbits.  Equations
(\ref{zeta_sol}) thus provide valid solutions for all Fourier modes of
$\chi^\alpha$ and $\zeta_\alpha$, except for the $k_r=0$ modes.
Given the fact (\ref{zeromodeSchw}) that $\hat F_\alpha(0,J)=0$, we see that a separate solution to Eqs.~(\ref{fcondFouri}) and (\ref{FcondFouri}) for the case $k_r=0$ is given by
\be\label{zerosol}
\hat\zeta_\alpha(0,J)=\left(\frac{\doe\omega^\beta}{\doe J_\alpha}\right)^{-1} \hat f^\beta(0,J),\qq H_1=0,
\ee
[with $\hat\chi^\alpha(0,J)$ unconstrained], provided that the matrix
$\doe\omega^\beta/\doe J_\alpha$ is invertible.  It follows from the
results of Ref.~\cite{isof} that $\doe\omega^\beta/\doe J_\alpha$ is
invertible for all stable bound geodesics, except for those
along the singular curve associated with isofrequency pairing of
Schwarzschild geodesics in the zoom-whirl regime (and those along the
separatrix defining the boundary of stable orbits); see Figure 1
of Ref.~\cite{isof}.
Thus, we have constructed a solution of Eqs.
(\ref{fcond}) and (\ref{Fcond}), and so the perturbed motion is
Hamiltonian and integrable.

\section{Conclusion}\label{sec:conclude}

We have shown that the local first-order conservative self-force
dynamics in the Schwarzschild spacetime is an integrable Hamiltonian
system, to linear order in the mass ratio, for generic stable bound
orbits outside the zoom-whirl regime (more specifically, for all
orbits to the right of the ``singular curve'' in Figure 1 of
Ref.~\cite{isof}).  The Hamiltonian system is defined by
Eqs.~(\ref{qJbarsystem}), with the coordinates $(\bar q^\alpha,\bar
J_\alpha)$ defined in terms of the geodesic action-angle coordinates
by Eqs.~(\ref{qJbar}), and with the Fourier modes of the functions
$\chi^\alpha$ and $\zeta_\alpha$ given by Eqs.~(\ref{zeta_sol}) and
(\ref{zerosol}).
The quantities $-\bar J_t$ and $\bar J_\phi$ are well-defined functions on the worldline phase space, which are perturbed versions of the geodesic energy $E$ and angular momentum $L_z$, which are instantaneously conserved to linear order under the conservative-self-forced motion.

Finally, we remark that the obstacles encountered by our construction
in the general Kerr case do not show that the system is not
Hamiltonian or integrable in Kerr.  As mentioned in the introduction,
there is a heuristic argument indicating that the dynamics should be
integrable and conservative in Kerr.

\acknowledgements

We thank Scott Hughes and Leo Stein for helpful conversations. This work was supported in part
by NSF grants PHY-1404105 and PHY-1068541.


\begin{thebibliography}{29}
\expandafter\ifx\csname natexlab\endcsname\relax\def\natexlab#1{#1}\fi
\expandafter\ifx\csname bibnamefont\endcsname\relax
  \def\bibnamefont#1{#1}\fi
\expandafter\ifx\csname bibfnamefont\endcsname\relax
  \def\bibfnamefont#1{#1}\fi
\expandafter\ifx\csname citenamefont\endcsname\relax
  \def\citenamefont#1{#1}\fi
\expandafter\ifx\csname url\endcsname\relax
  \def\url#1{\texttt{#1}}\fi
\expandafter\ifx\csname urlprefix\endcsname\relax\def\urlprefix{URL }\fi
\providecommand{\bibinfo}[2]{#2}
\providecommand{\eprint}[2][]{\url{#2}}

\bibitem[{\citenamefont{{Detweiler}}(2005)}]{Detweiler_rev05}
\bibinfo{author}{\bibfnamefont{S.}~\bibnamefont{{Detweiler}}},
  \bibinfo{journal}{Classical and Quantum Gravity}
  \textbf{\bibinfo{volume}{22}}, \bibinfo{pages}{681} (\bibinfo{year}{2005}),
  \eprint{gr-qc/0501004}.

\bibitem[{\citenamefont{Poisson et~al.}(2011)\citenamefont{Poisson, Pound, and
  Vega}}]{MPP}
\bibinfo{author}{\bibfnamefont{E.}~\bibnamefont{Poisson}},
  \bibinfo{author}{\bibfnamefont{A.}~\bibnamefont{Pound}}, \bibnamefont{and}
  \bibinfo{author}{\bibfnamefont{I.}~\bibnamefont{Vega}},
  \bibinfo{journal}{Living Reviews in Relativity}
  \textbf{\bibinfo{volume}{14}}, \bibinfo{pages}{7} (\bibinfo{year}{2011}),
  \eprint{1102.0529}.

\bibitem[{\citenamefont{Barack}(2009)}]{Barack_rev}
\bibinfo{author}{\bibfnamefont{L.}~\bibnamefont{Barack}},
  \bibinfo{journal}{Classical and Quantum Gravity}
  \textbf{\bibinfo{volume}{26}}, \bibinfo{pages}{213001}
  (\bibinfo{year}{2009}), \eprint{0908.1664}.

\bibitem[{\citenamefont{Barack}(2014)}]{Barack_rev14}
\bibinfo{author}{\bibfnamefont{L.}~\bibnamefont{Barack}}, in
  \emph{\bibinfo{booktitle}{General Relativity, Cosmology and Astrophysics}},
  edited by \bibinfo{editor}{\bibfnamefont{J.}~\bibnamefont{Bičák}}
  \bibnamefont{and} \bibinfo{editor}{\bibfnamefont{T.}~\bibnamefont{Ledvinka}}
  (\bibinfo{publisher}{Springer International Publishing},
  \bibinfo{year}{2014}), vol. \bibinfo{volume}{177} of
  \emph{\bibinfo{series}{Fundamental Theories of Physics}}, pp.
  \bibinfo{pages}{147--168}, ISBN \bibinfo{isbn}{978-3-319-06348-5},
  \urlprefix\url{http://dx.doi.org/10.1007/978-3-319-06349-2_6}.

\bibitem[{lis(2013)}]{lisa}
\bibinfo{journal}{The evolved Laser Interferometer Space Antenna/New
  Gravitational wave Observatory.}  (\bibinfo{year}{2013}),
  \urlprefix\url{http://www.elisa-ngo.org}.

\bibitem[{\citenamefont{{Le Tiec}}(2014)}]{LeTiec_rev}
\bibinfo{author}{\bibfnamefont{A.}~\bibnamefont{{Le Tiec}}},
  \bibinfo{journal}{International Journal of Modern Physics D}
  \textbf{\bibinfo{volume}{23}}, \bibinfo{eid}{1430022} (\bibinfo{year}{2014}),
  \eprint{1408.5505}.

\bibitem[{\citenamefont{Mino et~al.}(1997)\citenamefont{Mino, Sasaki, and
  Tanaka}}]{MiSaTa}
\bibinfo{author}{\bibfnamefont{Y.}~\bibnamefont{Mino}},
  \bibinfo{author}{\bibfnamefont{M.}~\bibnamefont{Sasaki}}, \bibnamefont{and}
  \bibinfo{author}{\bibfnamefont{T.}~\bibnamefont{Tanaka}},
  \bibinfo{journal}{Phys.~Rev.~D} \textbf{\bibinfo{volume}{55}},
  \bibinfo{pages}{3457} (\bibinfo{year}{1997}).

\bibitem[{\citenamefont{Quinn and Wald}(1997)}]{QuWa}
\bibinfo{author}{\bibfnamefont{T.~C.} \bibnamefont{Quinn}} \bibnamefont{and}
  \bibinfo{author}{\bibfnamefont{R.~M.} \bibnamefont{Wald}},
  \bibinfo{journal}{Phys.~Rev.~D} \textbf{\bibinfo{volume}{56}},
  \bibinfo{pages}{3381} (\bibinfo{year}{1997}).

\bibitem[{\citenamefont{{Gralla} and {Wald}}(2009)}]{GrallaWaldDGSF}
\bibinfo{author}{\bibfnamefont{S.~E.} \bibnamefont{{Gralla}}} \bibnamefont{and}
  \bibinfo{author}{\bibfnamefont{R.~M.} \bibnamefont{{Wald}}},
  \bibinfo{journal}{ArXiv e-prints}  (\bibinfo{year}{2009}),
  \eprint{0907.0414}.

\bibitem[{\citenamefont{{Gralla} and {Wald}}(2008)}]{rigor}
\bibinfo{author}{\bibfnamefont{S.~E.} \bibnamefont{{Gralla}}} \bibnamefont{and}
  \bibinfo{author}{\bibfnamefont{R.~M.} \bibnamefont{{Wald}}},
  \bibinfo{journal}{Classical and Quantum Gravity}
  \textbf{\bibinfo{volume}{25}}, \bibinfo{eid}{205009} (\bibinfo{year}{2008}),
  \eprint{0806.3293}.

\bibitem[{\citenamefont{{Rosenthal}}(2006)}]{Rosenthal_second-order}
\bibinfo{author}{\bibfnamefont{E.}~\bibnamefont{{Rosenthal}}},
  \bibinfo{journal}{\prd} \textbf{\bibinfo{volume}{74}}, \bibinfo{eid}{084018}
  (\bibinfo{year}{2006}), \eprint{gr-qc/0609069}.

\bibitem[{\citenamefont{{Gralla}}(2012)}]{Gralla_second-order}
\bibinfo{author}{\bibfnamefont{S.~E.} \bibnamefont{{Gralla}}},
  \bibinfo{journal}{\prd} \textbf{\bibinfo{volume}{85}}, \bibinfo{eid}{124011}
  (\bibinfo{year}{2012}), \eprint{1203.3189}.

\bibitem[{\citenamefont{{Pound}}(2012{\natexlab{a}})}]{Pound_second-order}
\bibinfo{author}{\bibfnamefont{A.}~\bibnamefont{{Pound}}},
  \bibinfo{journal}{Physical Review Letters} \textbf{\bibinfo{volume}{109}},
  \bibinfo{eid}{051101} (\bibinfo{year}{2012}{\natexlab{a}}),
  \eprint{1201.5089}.

\bibitem[{\citenamefont{{Pound}}(2012{\natexlab{b}})}]{Pound_nonlinear}
\bibinfo{author}{\bibfnamefont{A.}~\bibnamefont{{Pound}}},
  \bibinfo{journal}{\prd} \textbf{\bibinfo{volume}{86}}, \bibinfo{eid}{084019}
  (\bibinfo{year}{2012}{\natexlab{b}}), \eprint{1206.6538}.

\bibitem[{\citenamefont{Arnolʹd}(1989)}]{Arnold}
\bibinfo{author}{\bibfnamefont{V.~I.} \bibnamefont{Arnolʹd}},
  \emph{\bibinfo{title}{{Mathematical Methods of Classical Mechanics}}},
  Graduate Texts in Mathematics (\bibinfo{publisher}{Springer},
  \bibinfo{year}{1989}), ISBN \bibinfo{isbn}{9780387968902}.

\bibitem[{\citenamefont{Flanagan and Hinderer}(2012)}]{resonances}
\bibinfo{author}{\bibfnamefont{E.~E.} \bibnamefont{Flanagan}} \bibnamefont{and}
  \bibinfo{author}{\bibfnamefont{T.}~\bibnamefont{Hinderer}},
  \bibinfo{journal}{Physical Review Letters} \textbf{\bibinfo{volume}{109}},
  \bibinfo{pages}{71102} (\bibinfo{year}{2012}), \eprint{1009.4923}.

\bibitem[{\citenamefont{Hinderer and Flanagan}(2008)}]{two_timescale}
\bibinfo{author}{\bibfnamefont{T.}~\bibnamefont{Hinderer}} \bibnamefont{and}
  \bibinfo{author}{\bibfnamefont{E.~E.} \bibnamefont{Flanagan}},
  \bibinfo{journal}{Phys.~Rev.~D} \textbf{\bibinfo{volume}{78}},
  \bibinfo{pages}{64028} (\bibinfo{year}{2008}), \eprint{0805.3337}.

\bibitem[{\citenamefont{{Wu} and {Xie}}(2010)}]{Wu10}
\bibinfo{author}{\bibfnamefont{X.}~\bibnamefont{{Wu}}} \bibnamefont{and}
  \bibinfo{author}{\bibfnamefont{Y.}~\bibnamefont{{Xie}}},
  \bibinfo{journal}{\prd} \textbf{\bibinfo{volume}{81}}, \bibinfo{eid}{084045}
  (\bibinfo{year}{2010}), \eprint{1004.4549}.

\bibitem[{\citenamefont{Tichy and Flanagan}(2001)}]{wolfgang}
\bibinfo{author}{\bibfnamefont{W.}~\bibnamefont{Tichy}} \bibnamefont{and}
  \bibinfo{author}{\bibfnamefont{E.~E.} \bibnamefont{Flanagan}},
  \bibinfo{journal}{Classical and Quantum Gravity}
  \textbf{\bibinfo{volume}{18}}, \bibinfo{pages}{3995} (\bibinfo{year}{2001}).

\bibitem[{\citenamefont{{Isoyama} et~al.}(2014)\citenamefont{{Isoyama},
  {Barack}, {Dolan}, {Le Tiec}, {Nakano}, {Shah}, {Tanaka}, and
  {Warburton}}}]{Isoyama14}
\bibinfo{author}{\bibfnamefont{S.}~\bibnamefont{{Isoyama}}},
  \bibinfo{author}{\bibfnamefont{L.}~\bibnamefont{{Barack}}},
  \bibinfo{author}{\bibfnamefont{S.~R.} \bibnamefont{{Dolan}}},
  \bibinfo{author}{\bibfnamefont{A.}~\bibnamefont{{Le Tiec}}},
  \bibinfo{author}{\bibfnamefont{H.}~\bibnamefont{{Nakano}}},
  \bibinfo{author}{\bibfnamefont{A.~G.} \bibnamefont{{Shah}}},
  \bibinfo{author}{\bibfnamefont{T.}~\bibnamefont{{Tanaka}}}, \bibnamefont{and}
  \bibinfo{author}{\bibfnamefont{N.}~\bibnamefont{{Warburton}}},
  \bibinfo{journal}{ArXiv e-prints}  (\bibinfo{year}{2014}),
  \eprint{1404.6133}.

\bibitem[{\citenamefont{Damour}(2008)}]{intro_eob}
\bibinfo{author}{\bibfnamefont{T.}~\bibnamefont{Damour}},
  \bibinfo{journal}{International Journal of Modern Physics A}
  \textbf{\bibinfo{volume}{23}}, \bibinfo{pages}{1130} (\bibinfo{year}{2008}),
  ISSN \bibinfo{issn}{0217-751X}, \eprint{0802.4047}.

\bibitem[{\citenamefont{Buonanno et~al.}(2003)\citenamefont{Buonanno, Chen, and
  Vallisneri}}]{Buonanno03}
\bibinfo{author}{\bibfnamefont{A.}~\bibnamefont{Buonanno}},
  \bibinfo{author}{\bibfnamefont{Y.}~\bibnamefont{Chen}}, \bibnamefont{and}
  \bibinfo{author}{\bibfnamefont{M.}~\bibnamefont{Vallisneri}},
  \bibinfo{journal}{Phys. Rev. D} \textbf{\bibinfo{volume}{67}},
  \bibinfo{pages}{024016} (\bibinfo{year}{2003}).

\bibitem[{\citenamefont{Le~Tiec et~al.}(2012)\citenamefont{Le~Tiec, Blanchet,
  and Whiting}}]{LeTiec12}
\bibinfo{author}{\bibfnamefont{A.}~\bibnamefont{Le~Tiec}},
  \bibinfo{author}{\bibfnamefont{L.}~\bibnamefont{Blanchet}}, \bibnamefont{and}
  \bibinfo{author}{\bibfnamefont{B.~F.} \bibnamefont{Whiting}},
  \bibinfo{journal}{Phys. Rev. D} \textbf{\bibinfo{volume}{85}},
  \bibinfo{pages}{064039} (\bibinfo{year}{2012}).

\bibitem[{\citenamefont{{Le Tiec} et~al.}(2012)\citenamefont{{Le Tiec},
  {Barausse}, and {Buonanno}}}]{binding_energy}
\bibinfo{author}{\bibfnamefont{A.}~\bibnamefont{{Le Tiec}}},
  \bibinfo{author}{\bibfnamefont{E.}~\bibnamefont{{Barausse}}},
  \bibnamefont{and}
  \bibinfo{author}{\bibfnamefont{A.}~\bibnamefont{{Buonanno}}},
  \bibinfo{journal}{Physical Review Letters} \textbf{\bibinfo{volume}{108}},
  \bibinfo{eid}{131103} (\bibinfo{year}{2012}), \eprint{1111.5609}.

\bibitem[{\citenamefont{{Barausse} et~al.}(2012)\citenamefont{{Barausse},
  {Buonanno}, and {Le Tiec}}}]{Barausse11}
\bibinfo{author}{\bibfnamefont{E.}~\bibnamefont{{Barausse}}},
  \bibinfo{author}{\bibfnamefont{A.}~\bibnamefont{{Buonanno}}},
  \bibnamefont{and} \bibinfo{author}{\bibfnamefont{A.}~\bibnamefont{{Le
  Tiec}}}, \bibinfo{journal}{\prd} \textbf{\bibinfo{volume}{85}},
  \bibinfo{eid}{064010} (\bibinfo{year}{2012}), \eprint{1111.5610}.

\bibitem[{\citenamefont{{Schmidt}}(2002)}]{Schmidt}
\bibinfo{author}{\bibfnamefont{W.}~\bibnamefont{{Schmidt}}},
  \bibinfo{journal}{Classical and Quantum Gravity}
  \textbf{\bibinfo{volume}{19}}, \bibinfo{pages}{2743} (\bibinfo{year}{2002}),
  \eprint{gr-qc/0202090}.

\bibitem[{\citenamefont{Mino}(2003)}]{Mino}
\bibinfo{author}{\bibfnamefont{Y.}~\bibnamefont{Mino}},
  \bibinfo{journal}{Phys.~Rev.~D} \textbf{\bibinfo{volume}{67}},
  \bibinfo{pages}{84027} (\bibinfo{year}{2003}).

\bibitem[{\citenamefont{Grossman et~al.}(2013)\citenamefont{Grossman, Levin,
  and Perez-Giz}}]{Grossman:2011im}
\bibinfo{author}{\bibfnamefont{R.}~\bibnamefont{Grossman}},
  \bibinfo{author}{\bibfnamefont{J.}~\bibnamefont{Levin}}, \bibnamefont{and}
  \bibinfo{author}{\bibfnamefont{G.}~\bibnamefont{Perez-Giz}},
  \bibinfo{journal}{Phys.Rev.} \textbf{\bibinfo{volume}{D88}},
  \bibinfo{pages}{023002} (\bibinfo{year}{2013}), \eprint{1108.1819}.

\bibitem[{\citenamefont{{Warburton} et~al.}(2013)\citenamefont{{Warburton},
  {Barack}, and {Sago}}}]{isof}
\bibinfo{author}{\bibfnamefont{N.}~\bibnamefont{{Warburton}}},
  \bibinfo{author}{\bibfnamefont{L.}~\bibnamefont{{Barack}}}, \bibnamefont{and}
  \bibinfo{author}{\bibfnamefont{N.}~\bibnamefont{{Sago}}},
  \bibinfo{journal}{\prd} \textbf{\bibinfo{volume}{87}}, \bibinfo{eid}{084012}
  (\bibinfo{year}{2013}), \eprint{1301.3918}.

\end{thebibliography}
\end{document}